# PHYSICS IN DISCRETE SPACES (C):
## AN INTERPRETATION OF QUANTUM ENTANGLEMENT

Pierre Peretto
Laboratoire de Physique et Modélisation des Milieux Condensés
CNRS LPMMC Grenoble (France)
peretto.pierre@neuf.fr

Abstract:
In this contribution we use the model of discrete spaces that we have put forward in former articles to give an interpretation to the phenomena of quantum entanglement and quantum states reduction that rests upon a new way of considering space and time.
Pacs numbers:03.65.Aa, 03.65.Ta, 03.65.Ud

## 1. INTRODUCTION

Some implications of quantum theory seem to go against common sense. Perhaps the most intriguing implications are the phenomena of entanglement and quantum states reduction.

Two particles (1) and (2) of a quantum system are intertwined (or entangled) if the quantum state properties of, say, particle (1) cannot be described without appealing to the quantum state properties of the other particle (2) independently of the distance between the particles, giving rise to the so-called EPR paradox [1]. Recent experiments carried by Anton Zeilinger et al. [2] show that the modification of the polarization properties of one of two entangled photons instantaneously modifies the polarization properties of the other at distances larger than 120km. This apparently violates the principle of relativistic causality.

Another violation of the same principle occurs with the phenomenon of quantum states reduction. The spherical wave of a photon, emitted by an extremely distant star, propagates in free space according to Maxwell equations. The wave seems to instantaneously collapse in one point when the photon is observed in a detector. However once observed the photon can restart a new propagation and the same reduction effect can again take place afterwards.

Entanglement and quantum states reduction pose the question of the nature of space-time and, therefore, cannot be explained without appealing to a convenient model of space. The model of discrete spaces that we put forward in previous contributions, [3] and [4], could provide a relevant interpretation of these two effects.

## 2. A REMINDER OF THE DISCRETE SPACE-TIME MODEL

For a comprehensive understanding of the present approach we will, first of all, point out the main features of the model of discrete space-time that we propose.

i)- The universe is fully made of a countable set of very simple physical systems , called cosmic (physical) bits $\alpha$, ($\alpha = 1,2,\cdots$) whose states $\sigma_\alpha$ are completely determined by one (mathematical) bit of information $\sigma_\alpha = \pm 1$. The cosmic bits interact through random binary and, much weaker, quaternary interactions. It is assumed that the (non measurable) size of cosmic bits is of the order of the Planck length that is $\cong 10^{-33}$ cm .

ii)- Under the influence of their interactions the cosmic bits get organized in systems, called world points, comprised of *n* cosmic bits. The size *l\** of a world point would be of the order of $l^* \cong 0.5 \times 10^{-20}$ cm [4]. *l\**, called the metric limit, is the smallest length one can measure. Below this limit both the quantum states of quantum theory and the metric tensors of general relativity loose their meanings. The polarization state $\phi_i$ of world point *i* ( $i = 1, 2, \cdots, N$ ) is described by a normalized 4-dimensional vector

$$\phi_i = \begin{pmatrix} \varphi_{i1} \\ \varphi_{i2} \\ \varphi_{i3} \\ \varphi_{i4} \end{pmatrix}$$

with

$$\phi_i^T \phi_i = \sum_{\mu=1,\dots,4} |\varphi_{i\mu}|^2 = 1$$

iii)- The Lagrangian $\Lambda(\psi)$ of a state $\psi = \{\phi_i\}$ is given by

$$\Lambda(\psi) = \psi^T (\Delta \otimes G) \psi \qquad (1)$$

where $\Delta$ is a random, square, symmetric, $N \times N$ matrix whose elements $\Delta_{ij}$ describe the interaction between world points *i* and *j*. The interaction between two world points *i* and *j* is a sum of $n \times n$ binary random variables. Its distribution is therefore Gaussian and is given by

$$P(\Delta_{ij}) = \frac{1}{\sqrt{2\pi n}} \exp\left(-\frac{|\Delta_{ij}|^2}{2\pi n^2}\right). \qquad (2)$$

*G* is a set of *N*, square, symmetric, $4 \times 4$ matrices, $G_i$, whose elements $G_{i,\mu\nu}$ ( $\mu, \nu = 1, \cdots, 4$ ) represent the interactions between the components $\varphi_{i\mu}$ and $\varphi_{i\nu}$ of the polarization state $\phi_i$ of world point *i*.

iv)- The possible states $\psi = \{\phi_i\}$ of the whole universe are obtained by minimizing the Lagrangian under the normalization constraint $\psi^T \psi = \sum_i \phi_i^T \phi_i = N$ . This gives the following eigenvalue equation

$$(\Delta \otimes G)\psi = \kappa \psi \qquad (3)$$

where the eigenvalue $\kappa$ is a Lagrange multiplier.

## 3. THE CONSTRUCTION OF SPACE AND TIME

The equation (3) is, in fact, a Klein Gordon equation. The connection between eq.(3) and Klein-Gordon equation is established in Appendix A. Here is the link between the discrete and continuous descriptions of space-time. It could be tempting to associate definite values of space *x* and time *t* to each world point *i* that is $x = x(i)$ , $t = t(i)$ but this is not so. Let us remember that the interactions $D_{ij}$ are random variables distributed along eq. (2). The largest contributions to the first order and second order derivatives (A2) and (A3) come from the very rare highest values of $|D_{ij}|$. If there is only one such $D_{ij}$ then *j* is considered a close neighbour of *i*. The relation $\sum_j D_{ij} = \overline{D} \cong 0$ however compels the non vanishing $D_{ij}$'s to form a whole set $\Omega(i)$: The neighbourhood of *i* is so to speak distributed over the set $\Omega(i)$ of world points *j* linked to a world point $i$ by $|D_{ij}| > \varepsilon$. In fact these details are automatically taken into account in the Klein Gordon (A6) whose solutions are plane waves and write

$$\psi = \exp\!\left(i(kx\text{-}\omega t)\right) \qquad (4)$$

The physical systems are no more entities that move in a predetermined space-time, the background disappears, but they are entities that determine space-time itself. For example a straight line is an object defined by a constant phase $C$ of (4) : $kx - \omega t = C$ and it would be more correct to say that a straight line is defined by a light beam rather than to say that a light beam follows a straight line. Likewise a pendulum is generally considered as a physical system that measures time. In our opinion it would be more convenient to consider the pendulum as a physical system whose repetitive states introduce the notion of time. These arguments are close to the ideas developed by Carlo Rovelli [5].

## 4. HISTORIES

An eigenstate $\psi = \{\phi_i\}$, a solution of eq. (3), determines the state of the physical system for all world points $i$, therefore for all their neighbouring sets $\Omega(i)$, and finally for all space-time coordinates $(x,t)$. A formal expression of $\psi(x,t)$ may possibly be derived by using a suitable theory. Then an experimentalist looks at his watch which gives him the value of a time, that we call his proper time $\tau$ (proper to the experimentalist). Then the experimentalist replaces the time $t$ in $\psi(x,t)$ by $t = \tau$ and verifies whether the physical system is in state $\psi(x,\tau)$ or not. If it is both theoretician and experimentalist gain some confidence in the relevance of the theory that led to $\psi(x,t)$. We note that $\psi(x,t)$ gives the state of the system for any time, as well for times $t < \tau$ in the past as for times $t > \tau$ in the future. Therefore $\psi(x,t)$ describes the whole history of the system. It is essential to understand that a history must be considered as a specific single entity, not as a stack of various spatial states exactly as a sphere is to be considered as a single specific object not as a stack of circular disks.

In general, the eigenspace associated with $\psi(x,t)$ is highly degenerate and one needs to fix some constraints so as to determine which state is actually realized. In a particular experiment the constraints are determined by the experimental set up. The constraint consists, for example, in given values of some interactions $D_{ij}$ and given polarization states $\phi_i$ of some world points. The realized history in experiment A $\psi_A(x_A,t_A)$ determines a specific space $x_A$ and time $t_A$ according to the considerations that we have developed so far. The experimentalist, however, may eventually modify the experimental constraints. He then carries out a new experiment B which determines a new history $\psi_B(x_B,t_B)$ and this history, in turn, determines a new specific space $x_B$ and time $t_B$. The question arises to understand how $\psi_A(x_A,t_A)$ transforms into $\psi_B(x_B,t_B)$ during the modification of experimental constraints. The dynamics of this transformation cannot be expressed in terms of a physical time $t$ except if $\psi_A$ and $\psi_B$ belong to the same eigenspace because $t$ is only defined inside the eigenstates and not between the eigenstates. The only sort of time that could play a role is the proper time $\tau$ of the experimentalist. The proper time $\tau$ however has no meaning inside the histories. Its modification has no consequence on histories whatsoever which means that any dynamics described in terms of $\tau$ looks instantaneous as far as histories (and physical time $t$) are concerned. Let us illustrate these arguments on a very simple model that of a falling body. The history of this system, in the classical limit, writes

$$\psi(z,t) = \delta\!\left(z - 1/2 g t^2\right)$$

We consider two types of histories (A and B) (Fig.1)

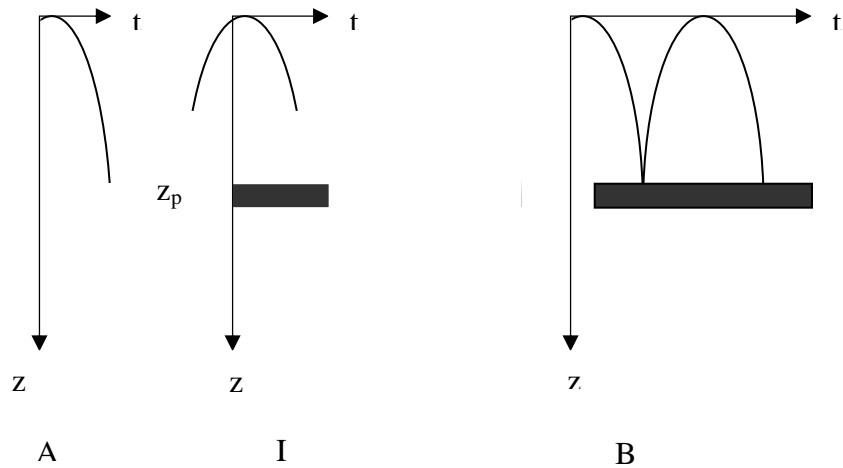

Fig.1-Histories of falling bodies:
A: The history of a free-falling body,
the parabola is the graph to be considered as a single entity
I: While the body is falling an obstacle is placed on its
trajectory at $z = z_p$

B: As soon as the obstacle is placed history A
is transformed into history B

i)- In experiment A there is no limitation to the fall. The history is a parabola.
ii)- In experiment B an obstacle is placed by the experimentalist at coordinate $z_p$ and the body bounces on the obstacle. The history, if damping effects are ignored, is periodic.
Let us assume that the experimentalist puts in place the obstacle at $z_p$ during the fall. As soon as the obstacle is in position at $z_p$ the history of the body changes from A to B and this modification, from the point of view of the falling body seems to be instantaneous. It is right to emphasize that the modifications of the experimental conditions are not instantaneous from the point of view of the experimentalist and takes a proper time duration $\Delta\tau$ but the details of the operation are completely ignored by the body provided that the operation is over when the body hits the obstacle.

## 5. ENTANGLEMENT

A typical entanglement experiment involves a pair of particles (1) and (2) emitted by a single source with zero total polarization, for example a pair of photons [6]. According to the formalism of quantum theory the state of a pair of entangled bosons may be written as

$$|1,2\rangle = \psi(1,2) = \frac{1}{\sqrt{2}}\left(\psi_{1\uparrow}\psi_{2\downarrow} + \psi_{1\downarrow}\psi_{2\uparrow}\right) = \frac{1}{\sqrt{2}}\left(\left|1\uparrow\rangle\right|2\downarrow\rangle + \left|1\downarrow\rangle\right|2\uparrow\rangle\right).$$

We do not know *a priori* the polarization state of any of the two particles but if a measurement shows that the polarization of (1) is, for example, $\uparrow$ then the polarization of (2) is necessarily $\downarrow$. For this observation to be carried out we position a polarizer $P$ on the path of particle (1). The state of the polarizer is either $P_\uparrow$ or $P_\downarrow$ and under the control of a first experimentalist E1. The polarizer may be seen as a projector operator. For example

$$P_\uparrow \;=\; \big|1\;\uparrow\;\big\rangle\big\langle 1\;\uparrow\;\big|$$

and

$$P_\uparrow|1,2\rangle = \big|1\uparrow\big\rangle\big\langle 1\uparrow\big|\big|1,2\big\rangle = \frac{1}{\sqrt{2}}\big|1,\uparrow\big\rangle\big|2\downarrow\big\rangle.$$

That is particle (1) is in state $\uparrow$ and particle (2) in state $\downarrow$ indeed.

We also position a polarization detector D on the trajectory of particle (2). The polarization is observed by a second experimentalist E2. The two experimentalists E1 and E2 have synchronized watches and therefore the same proper times $\tau$. Let us start the experiment with P in position $P_\downarrow$. The system follows a history A where the polarization of (2) is necessarily $\uparrow$ which is indeed observed by E2. At his proper time $\tau_1$ E1 positions the state of the polarizer along $P_\uparrow$. Then the history B of the system is instantaneously determined for all positions $x$ and for all physical times $t > \tau_1$. According to history B the polarization of (2) is $\downarrow$ even though, along the classical point of view, (1) has already passed the polarizer P. E2 observes that the polarization of (2) is $\downarrow$ indeed at his proper time $\tau_2$. This strange, non local, phenomenon is experimentally well documented and shows that the classical point of view is misleading. Moreover, since the transformation of histories looks instantaneous from the point of view of particles one may have $\tau_2 - \tau_1 < x_{12}/\mathrm{c}$ where $x_{12}$ is the distance between the experimentalist and c the speed of light, so possibly violating the principle of relativistic causality..

The following metaphor (which obviously is not a proof) may help. We consider that a movie is a history (in the sense defined above) in a 3-dimensional space made of two space dimensions and one time dimension. The two dimensional space is constituted by the set of images printed on the film and the one dimensional time space is formed by the successive notches which move the film forward. The time $t$ experienced inside the movie by the protagonists has nothing to do with the notches. The speed of the film, that is the speed of the gear that moves the notches, is controlled by the projectionist. The notch number determines the proper time $\tau$ of the projectionist. The projectionist may seen as an experimentalist and the script of the movie as a particular experimental set up that determines the history described in the movie. Let us assume that two movies are realized that are based upon two scripts that only differ by one detail. For example in the script of movie A the cat of the main character dies at the beginning of the story whereas he stays alive in the script of movie B. Let the projectionist starts the movie A. After a while the projectionist stops the projector on a certain notch and changes the film A for the film B. When the projector restarts the main character, in the world of the movie that evolves according to $t$, does not feel any time discontinuity but he his faced with a strange situation. He knows that the cat is dead and however he sees the cat alive exactly as the particle (2), who "knows" that $P$ is in state $P_\downarrow$, "sees" that $P$ is in state $P_\uparrow$.

## 6. QUANTUM STATES REDUCTION

The phenomenon of quantum states reduction can also be understood in terms of histories. We consider an experiment where a photon, emitted by a very distant sources S, possibly activates a detector D.

We have two sorts of experimental set ups.

i- In set up A, a photon source S, located at point $x_S$ in the three dimensional space, emits a photon in vacuum space.

We know that the history of the photon obeys a propagation equation

$$\frac{\hbar^2}{c^2}\frac{\partial^2 \psi}{\partial t^2} = \hbar^2 \Delta \psi$$

It is described therefore by a plane wave $\psi_k(x,t)$

$$\psi_k^A(x,t) \propto \exp\left(i\hbar\left(\omega_k t - kx\right)\right)$$

with $\omega_k = c|k|$. The vector $k$ fully characterizes a history and we may formalize $\psi_k(x,t)$ simply by a ket vector $|k\rangle$. A particular history is anisotropic but the whole set of histories with $|k|$ given, reconstructs the spherical electromagnetic wave generated by Maxwell equations.

ii- In set up B a detector D is added at a point $x_D$ in the three dimensional space.

We assume that D is a two-states system. Let $|d\rangle$ be the state of the detector. Either D is activated $|d\rangle = |1\rangle$ or D is silent $|d\rangle = |0\rangle$.

The plane wave $|k\rangle$ created by S does not couple with D except if the wave vector $k$ is strictly parallel to the vector $x_D - x_S$ which is very unlikely. When $k$ is parallel to $x_D - x_S$ the state $|k\rangle$ can couple with the state $|d\rangle$. The histories of the coupled system can be expressed in terms of product vectors $|k\rangle|d\rangle = |k,d\rangle$. They are eigenvectors of the operator $\Lambda^{(D)}$

$$\Lambda^{(D)} = e_S c_k^+ c_k + e_D c_D^+ c_D + e_{S\text{-}D}\left(c_D^+ c_S + c_S^+ c_D\right)$$

with $e_S = \hbar\omega_k$. $c_k^+$ ($c_k$) is an operator that creates (annihilates) a photon $|k\rangle$. Explicitly

$$\Lambda^{(D)} = \begin{pmatrix} e_S & e_{S\text{-}D} \\ e_{S\text{-}D} & e_D \end{pmatrix}$$

To the two eigenvectors correspond two sorts of histories. The two eigenvalues are given by

$$\kappa_\pm = \frac{1}{2}\left[e_S + e_D \pm \sqrt{\left(e_S + e_D\right)^2 - 4\left(e_S e_D - e_{S\text{-}D}^2\right)}\right]$$

and the two types of histories ($\psi_\pm^B$) by

$$\psi_\pm^B(k) = \frac{1}{\sqrt{e_{S\text{-}D}^2 + \left(\kappa_\pm - e_D\right)^2}}\begin{pmatrix} \kappa_\pm - e_D \\ e_{S\text{-}D} \end{pmatrix}$$

The probabilities for the detector D to be activated are

$$\left|\left\langle \psi_\pm^B \middle| 0,1\right\rangle\right|^2 = \frac{e_{S\text{-}D}^2}{e_{S\text{-}D}^2 + \left(\kappa_\pm - e_D\right)^2}$$

and for D to remain silent by

$$\left|\left\langle \psi_\pm^B \middle| k,0\right\rangle\right|^2 = \frac{\left(\kappa_\pm - e_D\right)^2}{e_{S\text{-}D}^2 + \left(\kappa_\pm - e_D\right)^2}$$

In the limit of zero coupling constant $e_{S\text{-}D} = 0$ one has $\left|\left\langle \psi_\pm^B \middle| 0,1\right\rangle\right|^2 = 0$ : D remains silent. An efficient detector is characterized by the condition $|e_{S\text{-}D}| >> |\kappa_\pm - e_D|$. Then $\left|\left\langle \psi_\pm^B \middle| 0,1\right\rangle\right|^2 = 1$ and both types of histories end with an activated detector $|d\rangle = 1$ .

From the point of view of the observer a particle has effectively hit the detector D although no localized particle is involved in the process

## 7. DISCUSSION AND CONCLUSIONS

The reality of quantum entanglement and quantum states reduction forces the physicists to admit that quantum theory is non-local. They are phenomena that both question the very nature of space-time.

To understand the mechanism underlying those surprising effects a convenient model of space-time is therefore necessary. In previous contributions we have put forward a possible model of space-time that could account for this mechanism.

The central idea is that neither time nor space has an ontological status and that they must rather be considered as constructions built along a blueprint provided by eigenstates of the Lagrangian of the system. The eigenstates, called histories, depend on experimental conditions. If the conditions are modified the eigenstates are also modified and space and time have to be reconstructed anew. If the experimental conditions are modified while the experiment is in progress the dynamics of the system proceeds as if the new conditions have existed from the start but this takes effect only after the modifications come into effect, so securing the principle of classical causality.

The experimental setups determine which history materializes. Two histories may be so different that no physical (natural) process can transform one into the other. For example we do not know any physical mechanism that spontaneously transforms a polarizer $P_\uparrow$ into a polarizer $P_\downarrow$. The experimentalist, however, has the power to build bridges between histories, and to change $P_\uparrow$ into $P_\downarrow$. The paradox is that the experimentalist himself belongs to nature.

To solve the paradox one must understand that the experimentalist has a wonderful machine at his disposal, namely his brain. The model of discrete universe that we propose is a sort of spin glass where binary entities interact through binary random interactions. We can consider that the brain is also a sort of spin glass where the neurons play the role of world points and the synapses the role of binary interactions [7]. In reality the structure of the human brain is much richer than the structure of space-time. There are of the order of $10^{10}$ synapses in a human brain. The connectivity between the neurons is similar to the connectivity in a 30 dimensional hypercube and, more than anything else, the connections between the neurons are not random but are the result of learning processes. We can consider that the possibility for human beings to make bridges between situations (histories) that would remain unrelated otherwise is an act of creation. Creation is a process that, so to speak, places the experimentalist outside the usual realm of natural phenomena and the actual observation of the entanglement phenomenon needs the intervention of a human mind.

APPENDIX A: EQUATION (3) IS A KLEIN GORDON EQUATION

The interaction matrix $\Delta$ can be factorized along $\Delta = D^T D$ where $D$ is a $N \times N$ upper triangular random matrix ($D_{i>j} = 0$), and $D^T$ the transpose lower triangular matrix of $D$, that is $D_{ji}^T = D_{ij}$. $D_{ij}$ is a random variable that also follows the distribution law (2). As such the eigenvalue equation (3) can be expanded along the components $\varphi_{i\mu}$ of the polarization state $\phi_i$

$$\sum_{j,k,\nu} D_{ij}^T G_{j,\mu\nu} D_{jk} \varphi_{k,\nu} = \kappa \varphi_{i\mu} \qquad (A1)$$

At this point the indices $i,j,..$ are only but ordinal numbers and have no physical meaning as far as space-time or other physical properties such as fields or particles, are concerned. $D$, however, may be seen as a discrete differential operator.

A vector field $\psi$ is a vector in the internal space of a world point and may be expanded on the polarization state components

$$\psi_{i\nu} = \sum_\nu C_{\mu\nu} \varphi_{i\mu}$$

where the $C_{\mu\nu}$'s are the coefficients of the expansion. One defines an increment $\delta\varphi_{i\mu}$ of $\varphi_{i\mu}$ along $\mu$ by

$$\delta\varphi_{i\mu} = \sum_j D_{ij} \varphi_{j\mu}$$

An increment of the $\nu$ component of vector field $\psi_i$ along dimension $\mu$ writes

$$\delta_\mu \psi_{i\nu} = C_{\mu\nu} \delta\varphi_{i\mu} = C_{\mu\nu} \sum_j D_{ij} \varphi_{j\mu}$$

$\delta$ is distributive with respect to additions of polarizations $\delta\left(\psi^1{}_i + \psi^2{}_i\right) = \delta\left(\psi^1{}_i\right) + \delta\left(\psi^2{}_i\right)$ and obeys the Leibnitz formula $\delta\left(\psi_i^1 \psi_i^2\right) = \psi_i^1 \delta\left(\psi_i^2\right) + \psi_i^2 \delta\left(\psi_i^1\right)$ the two properties of differential operators (see [3]).

The first order partial derivatives of the vector field components is then defined by

$$\partial_\mu \psi_{i\nu} = \frac{1}{l^*} \delta_\mu \psi_{i\nu} = \frac{C_{\mu\nu}}{l^*} \sum_j D_{ij} \varphi_{j\mu} \qquad (A2)$$

and the second order partial derivatives by

$$\partial_\mu^2 \psi_{i\nu} = \frac{C_{\mu\nu}}{l^{*2}} \sum_{jk} D_{ij}^T D_{jk} \varphi_{k\mu} \qquad (A3)$$

By using a basis where $G$ is diagonal ($G_{\mu\nu} = G_\mu \delta_{\mu\nu}$) and assuming that $G$ is site independent, the equation (4) becomes

$$\sum_{j,k} D_{ij}^T G_{j,\mu} D_{jk} \varphi_{k,\mu} = \kappa \varphi_{i\mu}$$

With (6) the eigenvalue equation (4) gives a set of four equations

$$G_\mu \partial_\mu^2 \psi_{i\nu} = C_{\mu\nu} / l^{*2} \; \kappa \psi_{i\mu} \qquad (A4)$$

By summing in both sides over index $\mu$, eq.(7) is written as

$$\sum_\mu G_\mu \partial_\mu^2 \psi_{i\nu} = \left(\kappa / l^{*2}\right) \sum_\mu C_{\mu\nu} \psi_{i\mu} = \left(\kappa / l^{*2}\right) \psi_{i\nu} \qquad (A5)$$

In vacuum the metric matrix $G$ is given by (see [3])

$$G = \begin{pmatrix} -1/c^2 & & & \\ & 1 & & \\ & & 1 & \\ & & & 1 \end{pmatrix}$$

and one defines the mass $m$ of the particle associated with field $\psi$ by $\kappa / l^{*2} = \left(mc^2 / \hbar\right)^2$. Finally one obtains

$$\left(\frac{1}{c^2}\frac{\partial^2}{\partial t^2} - \Delta + \left(\frac{mc^2}{\hbar}\right)^2\right)\psi_\nu(r,t) = 0 \qquad (A6)$$

that is a set of four Klein-Gordon equations.

Acknowledgment: I thank Dr Ana Cabral for her careful reading of this text.